# Understanding the Performance of Ceph Block Storage for Hyper-Converged Cloud with All Flash Storage

Moo-Ryong Ra

mra@research.att.com

AT&T Labs – Research

## Abstract

Hyper-converged cloud refers to an architecture that an operator runs compute and storage services on the same set of physical servers. Although the hyper-converged design comes with a number of benefits, it makes crucial operational tasks, such as capacity planning and cost analysis, fairly complicated. The problem becomes more onerous if we consider a complex distributed system, such as Ceph, for the cloud with the proliferation of SSD drives. In this paper, we aim to answer some of these questions based on comprehensive microbenchmarks, and consequently better understand the behavior of Ceph in a hyper-converged cloud with all-flash storage. We reported our findings based on the study, devised a cost model and compared the cost of hyper-converged architecture with dedicated storage architecture. Additionally we summarized our experience based on the interactions with many teams at AT&T in the past couple of years.

## 1 Introduction

Hyper-converged cloud refers to an architecture where an operator runs compute and storage services on the same physical servers. The design is often motivated by the fact that infrastructure resources are vastly underutilized most of the time. Thus, by consolidating compute and storage resources together, the operator can reduce the needs for purchasing more hardware significantly, thereby saving CAPEX/OPEX[1] (Sec. 2). Although hyper-converged architecture brings a number of benefits to the cloud infrastructure providers, it makes some essential operational tasks, e.g., capacity planning and cost modeling, more complex compared to the traditional architecture where we run storage systems on separate physical servers (as appliances). Recent adoption of high performance SSD drives make these tasks even more difficult because performance bottleneck will be shifted to shared resources, such as compute and/or networking elements rather than the disk drives [15, 3, 17, 16].

In this paper, we make a case for building a hyper-converged cloud with all flash storage and provide practical answers to the raised concerns. Specifically, we presented a set of microbenchmark data to aid capacity planning task (Sec. 4), shared deployment issues that we have faced in the past couple of years (Sec. 5), and devised a cost model (Sec. 6). We elaborate each of the contributions as follows.

First, in the measurement study on a carefully designed cluster (Sec. 3), we aim to understand the followings: a) where the performance bottleneck lies in per storage configuration and workload, b) how much system resources are consumed and c) what type of SSD we will need to better optimize infrastructure cost. Some of our results reinforce the known observations in the storage systems community, e.g., the performance bottleneck is shifted from disk to cpu, network or storage daemons. Others are more specific to the configuration that we used for the measurement. For instance, we found that, because of the consistency guarantees provided by Ceph and storage pool configuration, the system resources consumed by write and read IOs are largely asymmetric (up to 9x difference). Second, we share our experience on bringing a new software-based storage solution (in contrast to appliances) to production datacenters (Sec. 5). We discuss many dimensions which need to be addressed before it goes live. It is often the case that each of these items took a significant amount of time to be resolved. Lastly, we develop a cost model for comparing 5 year total cost of ownership (TCO) between traditional dedicated storage architecture and hyper-converged architecture. Our calculation based on the model demonstrated a potential benefit of the hyper-converged architecture (up to ∼2x if # of datacenters are close to 500).

The rest of the paper is organized as follows. Sec. 2 discusses a hyper-converged architecture and describes a software stack used for our study. Sec. 3 and Sec. 4 elaborates the details of the conducted measurement study. We shares our experience on deployment issues in Sec. 5. Then, we present a cost model in Sec. 6 and conclude the paper.

## 2 Background and Scope

### 2.1 Hyper-Converged Cloud (HCC)

Recently hyper-converged cloud (HCC) architecture received quite a bit of attention in the industry [8, 13, 1,

---
[1]CAPital EXpense/OPerational EXpense

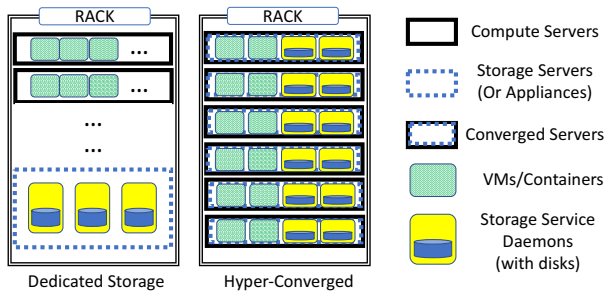

Figure 1: Dedicated Storage vs. Hyper-Converged Architecture

11]. The architecture allows infrastructure providers to run compute, storage, and networking elements in the same set of physical servers as illustrated in Fig. 1. In this architecture, the system resources such as cpu and memory will be shared not only among different tenant applications but also among infrastructure services, e.g., virtualization layer, storage systems and control plane (Openstack) functions. HCC architecture has several benefits to infrastructure providers. It enables uniform hardware deployment which makes lots of management decision much easier. Moreover, if we have fairly low infrastructure utilization, the benefit of the architecture will be magnified since we do not need to buy additional hardware just for storage space, and thereby reducing capital and operation cost (CAPEX/OPEX) significantly. We analyze the potential cost difference later in Sec. 6.

However, as implicitly discussed, one drawback of HCC architecture is that its compute-to-storage ratio is fixed. Sometimes separating compute from storage is important – so that each element can scale independently – if there is a significant skew towards either resource. To resolve this issue, some researchers examine a disaggregated deployment of storage services focused on a couple of storage-intensive applications [18]. However, this type of scenario is out of scope for this paper.

### 2.2 Openstack Cloud Operating Systems

We use Openstack [9] as a control plane for the HCC cloud. Openstack is an open source project which can be used to build a private cloud [10]. Its design separates control plane functions from data plane. Control plane functions are divided into several subcomponents. Examples of a sub-component could be a life cycle management service for VMs (*nova*), volumes (*cinder*), virtual networks (*neutron*), messaging among those services (*rabbitmq*) or resource accounting, i.e., vcpus, memory/storage space for VMs, etc. On the other hands, Openstack relies upon other open source projects and/or commercial software to implement its data plane. As for storage data plane, open source solutions, such as Linux iSCSI or Ceph, can be used to provide block storage to VMs or it can be implemented using vendor solutions as long as they implemented a driver for *cinder* service. In our study, we use Ceph and its block device (RBD) driver for *cinder* to provide VMs with block storage.

### 2.3 Ceph: Distributed Block Storage for the Cloud

We use Ceph [2, 20] to provide tenants and/or control plane services with a block storage interface. Ceph is an open source distributed storage system and widely adopted by the industry in recent several years [10]. It can be deployed on top of commodity servers and support three different storage APIs to the users – object, block, and file system. Ceph's storage server process is called Object Storage Daemon (*OSD*) and typically deployed one per hard disk drive and potentially more than one for SSD drive. The base layer of Ceph is an object store (*RADOS*). Block (*RBD*) and file system (*CephFS*) APIs are built on top of the RADOS layer. Ceph's object layer does not have a centralized component in its datapath. Instead of having a centralized metadata server or anything of that nature, the data objects coming into the system are evenly distributed to available OSDs by CRUSH algorithm [21]. To enforce a higher-level storage policy, Ceph has a notion of *storage pool* and *placement group* where users can control data redundancy, e.g., 2x/3x replication or erasure coding, compute-to-storage ratio and the rules for data placement.

### 2.4 Scope

In this paper, our primary interests are block storage interface for VMs in Openstack-based private clouds. The workloads running in VMs could be control plane components, such as *nova* or *cinder*, and/or tenant applications, such as VNFs, databases, virtualized CDN, etc. In our measurement study, we focused on block layer performance with different IO parameter combinations, rather than capturing specific characteristics of a certain application. This is because our focus is to broadly understand the behavior of Ceph in our target configuration with a wide range of applications in mind.

## 3 Cluster Setup for Capacity Planning

The main purpose of this study is to provide system architects with a full spectrum analysis of our storage configuration so that they can make well-informed decisions. We design a cluster to keep this goal in mind. To compare and contrast various options, we configured a 10 node cluster[2] and each node has 4x NVMe drives and 8x SAS SSDs. Additional details on HW configuration are articulated in Table 1.

---
[2]In this paper, we use the terms 'server' , 'node' and 'host' interchangeably.

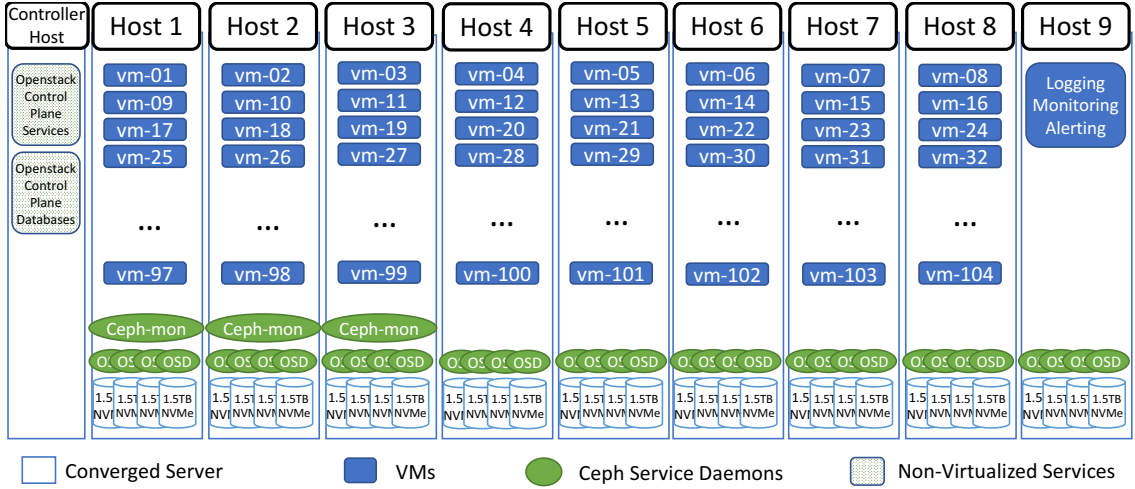

Figure 2: Mapping software components to physical servers: 1 OSD per NVMe drive

| Cluster Configuration | |
|---|---|
| Dell PowerEdge R730xd | 10x |
| Dell S6100 ToR switch | 1x |
| Server Specification | |
| Intel Xeon CPU E5-2695 v4 @ 2.10GHz | 2x |
| (total # of cores: 36, total # of vcpus: 72) | |
| Memory | 256 GB |
| NIC: Mellanox ConnectX-4 Lx 2x 25 GbE | 1x |
| Dell Express Flash NVMe SM1715 1.6TB (PCIe) | 4x |
| Toshiba PX04SRB096 900GB (SAS) | 8x |

Table 1: Cluster HW configuration

| Control plane and storage | |
|---|---|
| Ubuntu 16.04.02 LTS | all 10 nodes |
| Openstack Newton | 1 node |
| Ceph Storage | 9 nodes |
| Storage | |
| Ceph | Jewel 10.2.7 with filestore |
| config with NVMe SSDs | 36/72/108 OSDs |
| config with SAS SSDs | 72 OSDs (8 per host) |
| Performance monitoring | |
| Collector: Telegraf [12] | all 10 nodes |
| Database: InfluxDB [7] | 1 VM |
| Dashboard: Grafana [6] | 1 VM |

Table 2: SW configuration for our test cloud

Fig. 2 shows a high level mapping among software components and server nodes. We allocated the first node (*controller host*) to Openstack control plane services, such as nova, neutron, cinder, rabbitmq, etc., and used the rest 9 nodes (Host 1∼Host 9) for Ceph storage systems. The system resources of those 9 nodes are shared with tenant and monitoring VMs. Tenant VMs (fio-VMs) used the first 8 nodes (Host 1∼Host 8) and the last 1 node (Host 9) was used by monitoring VMs. For Ceph, the figure reflects a scenario where we deploy 1 OSD per NVMe SSD drive. In case of other OSD configurations we examined, e.g., 4 OSDs/NVMe or 1 OSD/SAS-SSD, the OSD/disk configuration part of the figure needs to be changed accordingly.

For monitoring the cluster, we use Telegraf [12] for data collection, InfluxDB [7] for time-series database and Grafana [6] for dashboard interface. All the graphs presented in the measurement section (Sec. 4) are based on the data collected through using the monitoring stack.

## 4 Microbenchmark: IO Performance and Resource Consumption

### 4.1 Data Collection

Adopting all flash storage in a hyper-converged cloud will provide high performance storage to tenants and control plane applications. However, it is crucial to understand what is the expected range of IO performance, how much system resources of the infrastructure are consumed, i.e., cpu, memory, disk and network, in order to operate the cloud.

To this end, we measured the performance using a popular I/O benchmark tool, FIO [4]. The tool provide a knob to control IO parameters to generate different types of synthetic IO workloads. Table 3 has a group of parameters used for this study. We tried all possible combinations of OSD configurations and fio parameters in the table, repeated at least 3 times for each combination and used averaged values for drawing graphs and tables. All data is collected using the monitoring stack described in Sec. 3. We analyze the dataset and make several interesting observations.

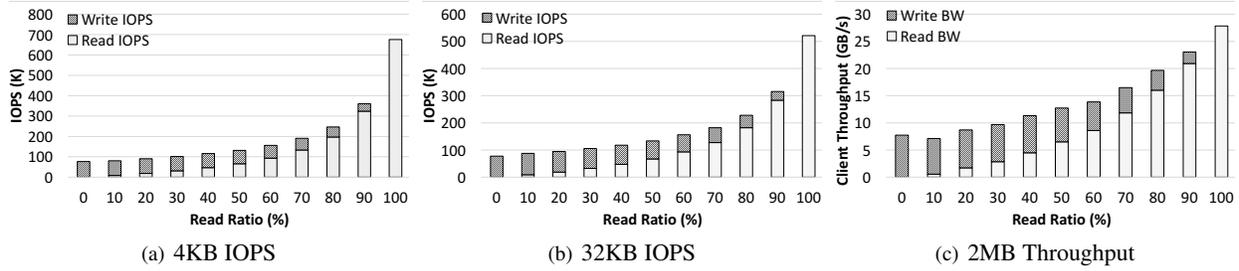

(a) 4KB IOPS  (b) 32KB IOPS  (c) 2MB Throughput

Figure 3: Baseline: max IOPS per block size (measured at the client)

| R/W IO Size | 0/100 | 10/90 | 20/80 | 30/70 | 40/60 | 50/50 | 60/40 | 70/30 | 80/20 | 90/10 | 100/0 |
|---|---|---|---|---|---|---|---|---|---|---|---|
| 4KB | 76446 (0/76446) | 80096 (8040/72056) | 90265 (18062/72203) | 101639 (30769/70870) | 116278 (46799/69479) | 131478 (65838/65640) | 155899 (93344/62555) | 190796 (133636/57160) | 246366 (196779/49587) | 360119 (323685/36434) | 675363 (675363/0) |
| 8KB | 90016 (0/90016) | 98655 (9952/88703) | 107512 (21585/85927) | 120117 (36378/83739) | 134618 (54104/80514) | 152388 (76217/76171) | 178022 (106722/71300) | 212649 (148938/63711) | 267051 (213463/53588) | 372267 (334410/37857) | 648848 (648848/0) |
| 16KB | 85830 (0/85830) | 94727 (9540/85187) | 103576 (20802/82774) | 114975 (34860/80115) | 128140 (51541/76599) | 145269 (72806/72463) | 168311 (100867/67444) | 201915 (141346/60569) | 254108 (202835/51273) | 355284 (319258/36026) | 610904 (610904/0) |
| 32KB | 77526 (0/77526) | 87644 (8781/78863) | 94802 (18912/75890) | 105614 (32024/73590) | 117408 (47276/70132) | 133676 (66931/66745) | 155924 (93450/62474) | 181889 (127194/54695) | 227686 (181798/45888) | 315226 (283331/31895) | 521175 (521175/0) |
| 64KB | 56734 (0/56734) | 68929 (6951/61978) | 76382 (15339/61043) | 84500 (25545/58955) | 96433 (38771/57662) | 108769 (54386/54383) | 125334 (75139/50195) | 150112 (105061/45051) | 185703 (148406/37297) | 245568 (220717/24851) | 375809 (375809/0) |
| 128KB | 38293 (0/38293) | 46261 (4627/41634) | 52061 (10405/41656) | 55709 (16782/38927) | 65286 (26236/39050) | 71783 (35853/35930) | 89174 (53461/35713) | 106881 (74866/32015) | 128052 (102306/25746) | 164965 (148253/16712) | 198500 (198500/0) |
| 256KB | 23007 (0/23007) | 27288 (2668/24620) | 28407 (5643/22764) | 32561 (9802/22759) | 36807 (14834/21973) | 42236 (21103/21133) | 50018 (29924/20094) | 60177 (42136/18041) | 70954 (56596/14358) | 83124 (74644/8480) | 100970 (100970/0) |
| 512KB | 11444 (0/11444) | 14370 (1372/12998) | 16318 (3135/13183) | 17109 (4978/12131) | 20603 (8132/12471) | 24404 (12165/12239) | 26096 (15717/10379) | 33210 (23192/10018) | 37113 (29581/7532) | 43383 (39074/4309) | 49534 (49534/0) |
| 1MB | 6312 (0/6312) | 7529 (731/6798) | 8303 (1503/6800) | 10434 (3060/7374) | 10697 (4250/6447) | 12225 (6125/6100) | 14016 (8459/5557) | 15400 (10791/4609) | 18086 (14483/3603) | 21873 (19726/2147) | 24977 (24977/0) |
| 2MB | 3868 (0/3868) | 3565 (278/3287) | 4342 (859/3483) | 4839 (1420/3419) | 5655 (2239/3416) | 6358 (3253/3105) | 6922 (4288/2634) | 8235 (5911/2324) | 9825 (8006/1819) | 11522 (10456/1066) | 13921 (13921/0) |
| 4MB | 1981 (0/1981) | 1783 (39/1744) | 2053 (416/1637) | 2679 (863/1816) | 3046 (1152/1894) | 3401 (1759/1642) | 4055 (2533/1522) | 4547 (3315/1232) | 4881 (4131/750) | 5502 (5245/257) | 6564 (6564/0) |

Figure 4: Max IOPS for each mixed workload configuration with different block size

| Test VMs & Ceph configurations | |
|---|---|
| VM flavor | 4 vcpus, 2GB memory |
| # of client VMs | 8∼104 (1∼13 per host) |
| Pool configuration | 3x replication with host separation. |
| OSD configuration | 1/2/4 OSDs/NVMe, 1 OSD/SAS-SSD |
| Benchmark Details | |
| Tool | fio-2.2.10 |
| IO engine / IO mode | libaio / direct IO |
| Block size (KB) | 4/8/16/32/ ... /1024/2048/4096 |
| Read/Write ratio | 0/100, 10/90, 20/80 .. , 90/10, 100/0 |
| IODepth | 8, 32 |
| Ramp-up time | 15 seconds |
| # of jobs per VM | 8 |
| Runtime | 5 minuites |
| IO Size (Filesize) | 100GB |

Table 3: Test configs and IO parameters used for our study

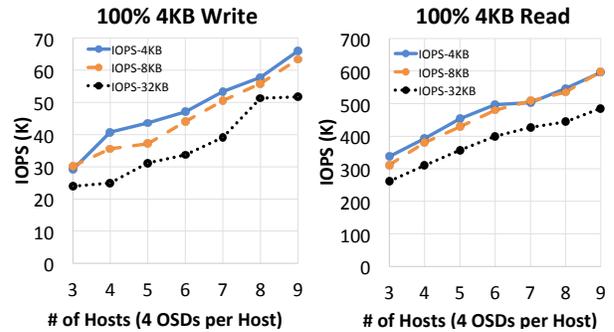

Figure 5: Scaling performance with # of hosts

### 4.2 Baseline: Performance and Scalability

The first obvious step for capacity/performance planning is to understand expected performance range under probable deployment scenarios.

To form a baseline, we first configured Ceph cluster so that it uses 8x SAS SSDs per node, i.e., 72 OSDs in total across 9 nodes, and ran a series of FIO benchmarks with different block size and read/write ratio. Fig. 3 shows IOPS numbers for different read/write mix for 4KB workload and Fig. 4 has a complete information including all the block sizes that we used for this measurement. Besides the presented performance numbers, we made a couple of observations. First, we can easily identify that, in general, *write operations are more expensive than read operations, especially for smaller block sizes*. This behavior is somewhat expected. For
4

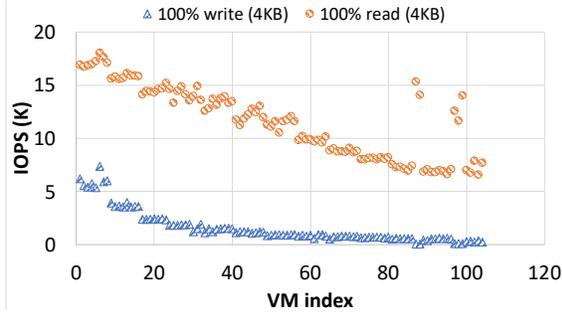

(a) Max IOPS per VM (bs=4KB)

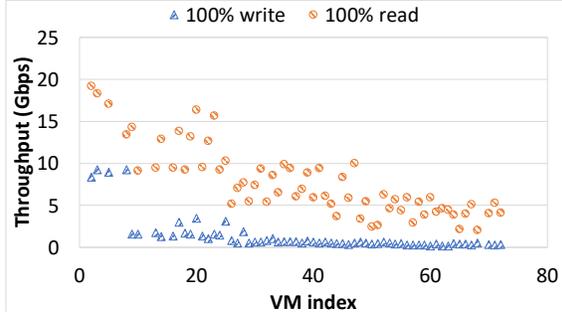

(b) Max Gbps per VM (bs=2MB)

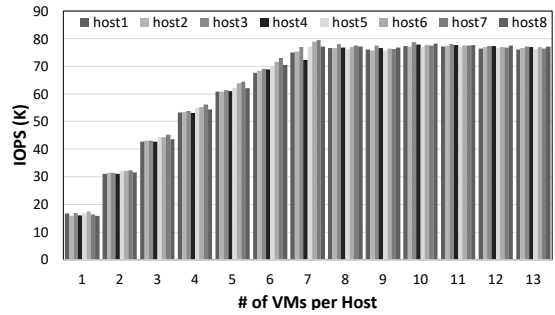

(c) Max IOPS per Host (bs=4KB)

Figure 6: Shifted Performance Bottleneck: per-VM or per-Host IOPS/Throughput limit exist

write operation, Ceph acknowledged to the client only when it fully writes all the required copies, i.e., 3 times in our case since we are using 3x replication. For read, the client will communicate with only one storage server (i.e., a primary OSD) and return to the client. As briefly mentioned, one interesting fact was that the difference between the required resources for read/write becomes larger as we use smaller block sizes. For example, if we calculate the ratio for 4KB IOs, write operations are ∼9x more expensive than read operations. Whereas, the ratio becomes closer to the expected number, i.e., 3x, if we increase the block size to 4MB. Second, intuitively IOPS would be inversely proportional to block sizes. However, it was only partially true for smaller block sizes in our measurement results. The max-IOPS numbers were only marginally decreasing from 4KB to 32KB range. What this means is that, if at all possible, cloud-friendly applications could utilize this fact to better utilize available network, while maintaining low response time from the underlying storage systems. Lastly, scalability is an important factor for planning the future of the storage cluster. Fig. 5 shows the results. In this test, we deploy 4 OSDs per host and each OSD is backed by an NVMe SSD drive. We started from 3 hosts and gradually increased # of hosts (nodes) one at a time. Overall IOPS scales linearly along with # of hosts. For 4KB workload, we can expect 6∼7K IOPS for write and ∼40K IOPS for read will be additionally supported by the cluster.

### 4.3 Performance Bottleneck

Traditionally, rotational medium (HDDs) typically form a performance bottleneck of a storage system. This obervation is no longer true with all flash storage. As a concrete example, Fig. 6 illustrates that, in our setup, a VM or a host can become a performance bottleneck instead of disk drives. Fig. 6(a) and Fig. 6(b) are from the experiments based on 72 client VMs, i.e., 9 VMs per host. For each VM index, we plot the maximum IOPS/throughput that the VM has ever shown during the experiments conducted in Sec. 4.2. In the experiments with 4KB workload, per-VM IOPS limit was around 18K IOPS. For 2MB workload, the limit per VM was around ∼20Gbps for read and ∼10Gbps for write. In Fig. 6(c), we increase # of VMs per host one at a time and plot per-host IOPS to see if the performance is saturated or not. As we can see in the figure, the bottleneck was forming as we reach ∼80K IOPS per host.

Combining Fig. 6 with the baseline results, we made an observation that *the performance bottleneck is often shifted from disks to somewhere else* in HCC with all-flash storage. Now the disks are no longer a bottleneck in a virtualized storage path. Apparently this observation is not new but fortifies the observation made by many researchers in the storage community [15, 3, 17, 16]. In our test cluster, we observed that the performance bottleneck can be formed in one of the following components: a) storage servers (OSDs) when the workload is mainly composed of write operations with small block sizes. b) network when the workload has a large block size. c) compute resources if a high level storage policy forces us to limit the storage systems' compute resource.

### 4.4 CPU Consumption

In HCC, cpu resources are shared among tenant VMs and storage systems. Since the capacity planning process needs to take the cpu resources into account in order to properly allocate them to tenant VMs, we need to effectively control the cpu consumption of underlying storage

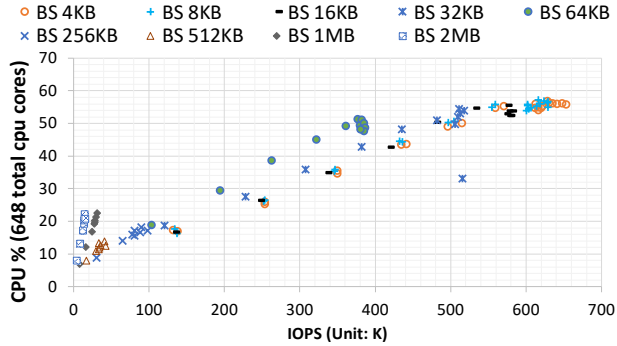

(a) 100% read CPU load, 1 OSD/NVMe

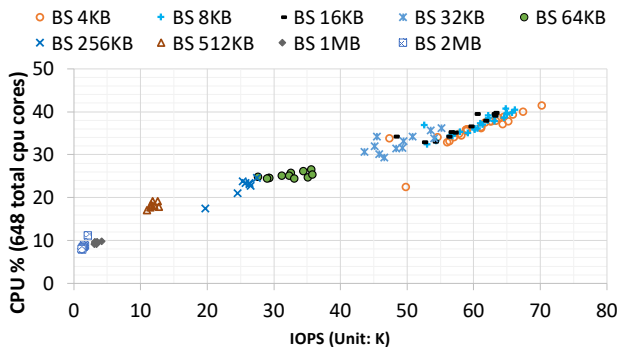

(b) 100% write CPU load, 1 OSD/NVMe

Figure 7: For 100% read/write workload, CPU load tends to be proportional to IOPS

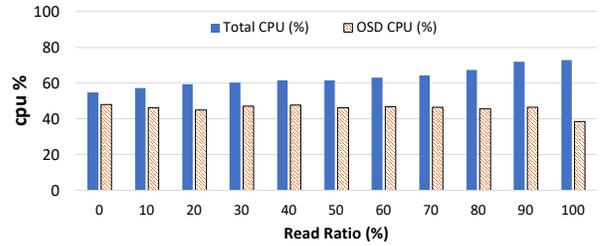

(a) Total vs. OSD CPU consumption (4KB)

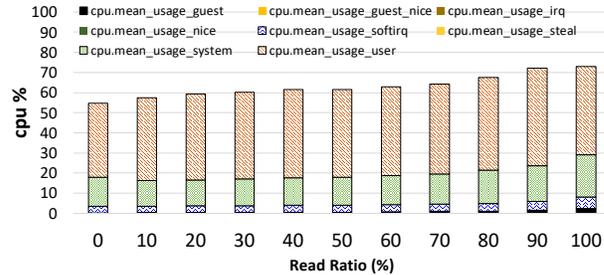

(b) Breakdown of CPU consumption per OS mode

Figure 8: Anatomy of CPU Consumption

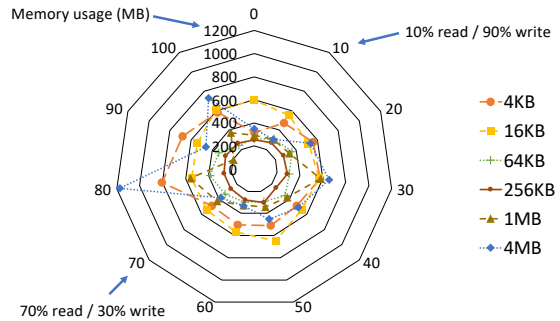

Figure 9: OSD's active memory consumption

systems. Fig. 7(a) and Fig. 7(b) show the impact of IOPS to total cpu consumption (In our setting, the total # of cores are 648 because 9 hosts are involved and each host has 72 cores (18 physical x 2 threads x 2 sockets). Thus, 50% of cpu consumption means that 324 cores out of 648 cores are fully utilized. According to the results, it is reasonable to say that the cpu load is proportional to IOPS. For read operations, the same observation holds but the slopes of larger block sizes (i.e., 64KB and larger) are a little deviated from the rest of the data points. However, for write, data points from all block sizes are relatively well-aligned one another.

Going one step further, we plot Fig. 8 to show two informations – a) total cpu consumption vs. OSD's cpu consumption for the 4KB workloads with different r/w ratios (left one) and b) the breakdown of total cpu consumption. For a), more than 90% of total cpu consumption went to OSD processes for 100% write. However, only about 50% is consumed by OSD for the 100% read case. What this means is that Ceph OSDs are doing a lot more computations for write operations compared to read operations. For b), most cpu cycles are spent by 4 os modes: user mode, system mode, softirq and guest mode. Among them, user mode takes the most ranging 61%~72% of total cpu consumption. Guest mode cpu consumption reaches 2.5% of total cpu capacity for 100% read workload, SoftIRQ (handling interrupts for network traffic and block IOs) consumes 2.5~5.6%, which is equivalent to consuming 16~36 cpu cores in total (or 2~4 cpu cores per host).

### 4.5 Memory footprint of storage servers (OSDs)

Fig. 9 illustrates average memory footprint (resident set size) of an OSD process during all experimental runs. The memory consumption per OSD was ranging from 150MB to 800MB (except one outlier reaching to 1.2GB). Since memory footprint is greatly affected by precedent condition, it is non-trivial to understand its exact behavior. Nonetheless, for planning purpose, it is still meaningful to figure out the range of active memory usage so that we can allocate remaining memory space to tenant VMs, for example.

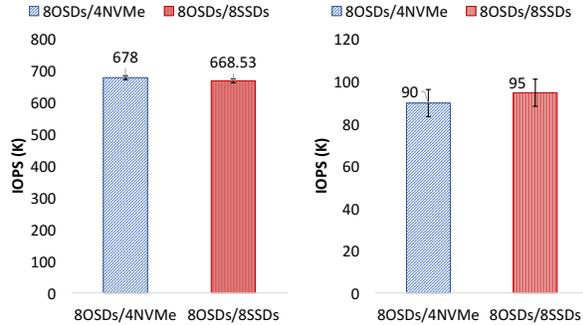

Figure 10: NVMe SSDs vs. SAS SSDs: max IOPS (4KB)

### 4.6 Do we need NVMe SSDs for Ceph in HCC?

In this subsection, we compare the impact of applying different type of SSDs to Ceph OSD processes to better understand cost-performance trade-offs. As for raw IO capability, high-end NVMe SSDs can provide an order of magnitude higher IOPS than regular SSDs, e.g., 700K vs. 70K, and 5∼10 times higher throughput, e.g., 3∼4 GB/s vs. 300∼500 MB/s. Then, the question is *if we use NVMe SSDs, how much benefit will we get compared to the configuration based on regular enterprise (SAS) SSDs?* To answer this question, we conducted two sets of experiments – one based on SAS SSDs and the other based on NVMe SSDs. For fair comparison, we used the same number of OSDs (8 per host). Fig. 10 shows the result, which is essentially saying that there is no performance difference between those two configurations. In other words, with today's Ceph storage solution under HCC architecture, it is not yet meaningful to have such a high performance drive, i.e., NVMe SSDs, given that NVMe SSDs is still significantly more expensive than regular (SATA/SAS) SSDs.

## 5 Road to Production

In this section, we share our experience in the past couple of years at a high level and discuss important items that we had to address before we bring a new storage solution to our production datacenters.

### 5.1 Reflecting the reality: sharing hardware and creating a path from a precedent condition

In an appliance-based datacenter, HCC architecture with a distributed storage system (such as Ceph) has a good value proposition because we can simplify deployment plans and save cost. However, in many cases, we need to consider already deployed sites, so-called "brown field", which may not have resources that we need to fully utilize architectural benefits. Devising a path from existing condition can be very challenging. We introduce two examples that we actually faced in our production environment.

**Sharing hardware resources:** In a HCC architecture, system resources will be shared across different services, e.g., compute, storage, and/or control plane. Therefore, in terms of required hardware resources, there will be conflict of interests among running services with reasonably high probability. If we have such a conflict, e.g., requiring local disks, it is often necessary to develop a solution to properly support both needs. As an example, we had a networking service relying on a Cassandra database and the database owner wanted to use local disks. This was problematic since our proposed solution is supposed to manage all available local disks using another solution. Carving out some of the available disks for a specific use case will cause too much operational complexity and additional development efforts for operation and development team. To resolve this issue, we had to come up with a specialized storage pool configuration and support both requirements simultaneously.

**Creating a path forward:** An initial proposal can uniformly manage local disks available in compute servers. However, it turned out that local disks are not available in a target datacenter. As a result, we had to discuss with several stakeholders to figure out whether it is possible to purchase local disks for the datacenter, what the related policies are, and then was able to devise a plan going forward.

### 5.2 Operationalization

When it comes to production, several things need to be prepared before the solution is going live. Especially, the solution heading to production datacenters are new to the operations team, the importance of these tasks become even greater.

First of all, it is very important to have a team/group/company who can provide *technical support with a desirable level of SLA*. In a large telecommunication company such as AT&T, this requirement is mandatory before any solution goes to the production datacenters. Second, one needs to provide a clear way of *monitoring the infrastructure* so that operators can make a timely decision for growth or be informed in case of potential failures/anomalies. Third, a thorough *capacity planning* is necessary to properly prepare a future growth of the infrastructure. We believe that the information shared in Sec. 4 can be a good example of this task. Fourth, we were often asked the reliability of the data that will be stored in our proposed solution from various teams. The key performance indicators (KPIs) for this item include both service availability, e.g., system uptime (%), and data durability. For service availability, common ac-tion items were to eliminate a single point of failures in all essential components, e.g., storage service daemons, hardware (physical servers, switches, etc.) and human errors. For the concerns around data durability, modern distributed storage systems provide a way to implement a storage policy, e.g., a storage pool configuration where we can set a replication factor/erasure coding, backup policy, etc. The last puzzle piece in this category is to have a well-established ***backup and restore plan***, which will help operators escape from some catastrophic failures. Fifth, all facets of ***security*** aspects need to be carefully examined so that we can quickly detect and confine potential problems.

| $HW_p$ | Annual HW purchase cost |
|---|---|
| $HW_s$ | Annual HW support cost |
| $SW_s$ | Annual software support cost |
| $PW$ | Annual power/cooling cost per datacenter |
| $N_{dc}$ | # of datacenters |
| $DEV$ | Annual development/certification cost |
| $FTE_{op}$ | Annual DevOps cost for operation |

Table 4: Variables for cost comparison

Decisions for some of the discussed items directly affect cost, e.g., replication factor, implementing policies for backup/restore and security practices. Not surprisingly, if we make the infrastructure more reliable and more secure, the cost will increase accordingly.

# 6 Cost Analysis

In a corporate environment, cost is important. Justifying new systems architecture from the perspective of cost is always very important. However, in many cases, it is difficult to make an apple-to-apple comparison of different storage solutions. It is often the case that a storage solution has its own assumptions behind provided $/GB numbers. Moreover, for more accurate comparison, we need to consider infrastructure cost as well, such as switches/rack cost, cooling, power source, etc. These costs will vary significantly according to the characteristics of storage solutions, such as density, average power consumption, temperature characteristics, etc. In this section, we develop a model that can mitigate the above-mentioned problems. Using the model, we quantify the cost of HCC architecture and compare it with an architecture where we use a dedicated servers for storage systems.

## 6.1 Cost model

To develop a relevant cost model that can be used for comparing various storage solutions, we take into account the following factors: HW purchase cost, HW support cost, SW support cost, networking HW purchase/support cost, power/cooling cost, human cost for development/certification/DevOps cost. The factors are sum-marized in Table 4. In terms of calculation, we used 5 year total cost of ownership (TCO). It can be calculated using the following equation.

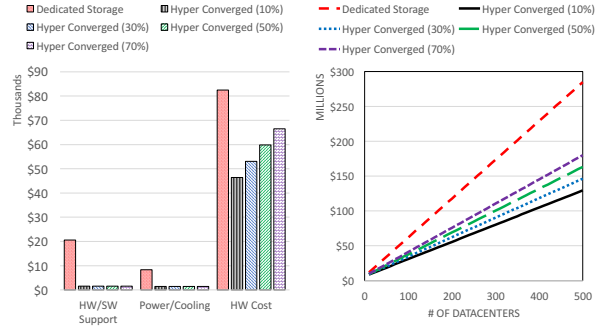

Figure 11: Cost components breakdown (left) and cost projection at scale (right): in this problem setting, hyper-converged design save significant cost in hardware cost, especially overall system utilization is low (left). Its benefit can be maximized as we have more number of datacenters (right).

$$\begin{aligned} Annual\ Cost &= \{HW_p + HW_s + SW_s + PW\} \cdot N_{dc} \\ &+ DEV/CERT + FTE_{op} \end{aligned}$$

In this equation, since the metric is 5 year TCO, $HW_p$ will be the total hw purchase cost divided by 5, etc.

## 6.2 Cost Projection at Scale

Our primary purpose for this exercise was to see the cost benefit of HCC architecture. For actual cost calculation, we utilize public sources [19, 5, 14] as well as our internal sources. We set our storage requirement per site as 50 TB and explore the case of up to 500 datacenters. These numbers are somewhat speculative but roughly aligned with a deployment plan related with AT&T's small-scale datacenters across the nation.

When we calculate cost for storage systems deployed in a HCC environment, we assume a certain percentage of the servers will be used for storage components since system resources are shared among all services running on the servers. As we observed in Sec. 4, accurate ratio will be dependent on performance requirements, i.e., if an IOPS requirement is higher, we will need to account more overhead. In this subsection, we did our estimation based on the HCC overhead ranging from 10% to 70%.

Fig. 11 shows the cost comparison result. In the left graph in the figure, we breakdown the cost component. Then, in the right graph, we examine how the cost projection changes as we increase the number of datacenters. In the comparison, HCC design demonstrated significant cost reduction because it reduces the needs for buying more servers. The cost benefit of HCC architecture is maximized as we have more # of datacenters. In addition, having a lesser IOPS requirement will help us reduce the cost as well in HCC architecture.

## 7 Conclusion

In this paper, we made a case for building hyper-converged cloud with all-flash storage based on Openstack and Ceph. We shared raw performance data as well as the lessons learned from our extensive measurement study. Some of them fortify the existing observation, i.e., performance bottleneck is shifted from the disks to somewhere else, while others help infrastructure designers make more cost-effective decisions, e.g., in our setting, NVMe SSDs are probably an excessive investment yet, etc. Moreover, we discussed a set of deployment issues when bringing new storage solutions to production datacenters. Lastly we described a cost model and projection at scale, comparing HCC architecture with dedicated counterparts.